%% file: root.tex
\documentclass[letterpaper, 10 pt, conference]{ieeeconf}  %

\usepackage[linesnumbered,ruled,vlined]{algorithm2e}

\IEEEoverridecommandlockouts                              %

\usepackage[dvipsnames]{xcolor}

\makeatletter
\let\NAT@parse\undefined
\makeatother

\usepackage{graphics} %
\usepackage{graphicx}
\usepackage{booktabs}
\usepackage{pifont}
\usepackage{tikz}     %

\usepackage{float}
\usepackage{subcaption}
\usepackage[font={footnotesize}]{caption}
\usepackage{epsfig} %
\usepackage{mathptmx} %
\usepackage{times} %
\usepackage{amsmath} %
\usepackage{amssymb}  %
\usepackage{multirow}
\usepackage{url}

\usepackage[bookmarks=true, colorlinks=true,linkcolor=magenta,citecolor=green!80!black]{hyperref} 
\usepackage[capitalise,noabbrev,nameinlink]{cleveref} %
\usepackage{mathrsfs}
\usepackage[normalem]{ulem}

\newcommand{\our}{\textsc{EPH}}

\newcommand{\citep}{\cite}
\newcommand{\citet}{\cite}

\title{\LARGE \bf
Ensembling Prioritized Hybrid Policies for Multi-agent Pathfinding
}

\author{Huijie Tang$^{*}$$^1$$^2$, Federico Berto$^{*}$$^1$$^2$, Jinkyoo Park$^1$$^2$
\thanks{$^*$Equal contributions}
\thanks
{$^1$Department of Industrial and Systems Engineering, KAIST, South Korea}
\thanks{$^2$OMELET}
\thanks{$^\ddagger$Authors are members of the AI4CO open research community.}
\thanks{Emails:
\{\textit{raylan.tang; fberto; jinkyoo.park}\}\textit{@kaist.ac.kr}}
}

\begin{document}

\maketitle
\thispagestyle{empty}
\pagestyle{empty}
\renewcommand{\labelenumi}{\textit{\roman{enumi})}} %

\begin{abstract}
Multi-Agent Reinforcement Learning (MARL) based Multi-Agent Path Finding (MAPF) has recently gained attention due to its efficiency and scalability. Several MARL-MAPF methods choose to use communication to enrich the information one agent can perceive. However, existing works still struggle in structured environments with high obstacle density and a high number of agents. To further improve the performance of the communication-based MARL-MAPF solvers, we propose a new method, Ensembling Prioritized Hybrid Policies (\our{}). We first propose a selective communication block to gather richer information for better agent coordination within multi-agent environments and train the model with a Q-learning-based algorithm. We further introduce three advanced inference strategies aimed at bolstering performance during the execution phase. First, we hybridize the neural policy with single-agent expert guidance for navigating conflict-free zones. Secondly, we propose Q value-based methods for prioritized resolution of conflicts as well as deadlock situations. Finally, we introduce a robust ensemble method that can efficiently collect the best out of multiple possible solutions. We empirically evaluate \our{} in complex multi-agent environments and demonstrate competitive performance against state-of-the-art neural methods for MAPF. We open-source our code at \url{https://github.com/ai4co/eph-mapf}.

\end{abstract}

\section{Introduction}
Multi-Agent Pathfinding (MAPF) involves finding collision-free paths for a group of agents while also aiming to minimize makespan or sum of costs \cite{stern2019multi}. MAPF has many practical applications, including warehouse robotics \cite{wurman2008coordinating}, aviation \cite{morris2016planning}, and digital gaming \cite{ma2017feasibility} scenarios. However, the challenge intensifies with the realization that optimal solutions for MAPF are NP-hard, involving solving large-scale constraint satisfaction problems in combinatorial spaces \citep{yu2013structure,banfi2017intractability}. Classical algorithms to solve the MAPF problems such as heuristic centralized solvers typically invoke search-based algorithms, e.g., Conflict-Based Search (CBS) \cite{sharon2015conflict}, as well as its improved versions ECBS \cite{barer2014suboptimal} and EECBS \cite{li2021eecbs}. However, as the number of agents increases, many heuristic solvers struggle to scale because of the high complexity of considering all agents in a centralized manner at once. In addition, in real-world applications, new tasks frequently occur during the execution of the initially planned paths \cite{li2021lifelong}. Such centralized heuristics solvers have to re-plan paths for all the agents after seeing new tasks, making it burdensome to apply to real-world applications.

Multi-Agent Reinforcement Learning (MARL) based approaches \cite{sartoretti2019primal, damani2021primal, ma2021learning, ma2021distributed} offer another way to solve the MAPF problem. Instead of treating MAPF as a centralized problem, they solve MAPF by regarding it as a sequential decision-making problem. One of the seminal MARL-based approaches to MAPF is PRIMAL\cite{sartoretti2019primal}. In PRIMAL, the authors propose to learn a fully decentralized policy with partial agent-wise observation; the model is trained with MARL and imitation learning (IL),  However, PRIMAL struggles to scale to high-obstacle-density environments with many agents. Also, PRIMAL assumes that agents move sequentially in each time step, which differs from real-world scenarios where agents move simultaneously in each time step. The improved version of PRIMAL, PRIMAL2 \cite{damani2021primal}, also trains with MARL and IL. PRIMAL2 assumes the \textit{disappear at target} \cite{stern2019multi} scenario. This assumption makes the problem easier because it means lower obstacle density when agents reach their goals and disappear, whereas in the \textit{stay at target} \cite{stern2019multi} setting, agents constitute obstacles since they do not disappear after reaching their goals. Thus, the problem is much harder, especially when many agents exist. We adopt \textit{stay at target} in this work.

Apart from the abovementioned MARL-based solvers \cite{damani2021primal,sartoretti2019primal} that consider environments that do not allow communication, another work direction is MARL-based solvers with communication between agents when communication is possible, such as DHC \cite{ma2021distributed} and DCC \cite{ma2021learning}. DHC is trained with Deep Q-learning \cite{mnih2013playing} and adopts a similar setting as PRIMAL, where each agent has a partial observation. However, DHC is different in that it introduces \textit{Graph Convolutional Communication}, which enables agents to communicate with other nearing agents for a better understanding of the environment, thus improving performance. Its improved version, DCC, improves communication by not learning broadcast but selective communication, thus reducing communication overhead. Recent MARL-based  MAPF works like SCRIMP \cite{wang2023scrimp} and SACHA \cite{lin2023sacha} also use communication mechanisms. For example, in SCRIMP, global communication is used to gather information from all agents in the environment, but this might be inefficient and burdensome in real-world applications.

\textit{Contributions.} In this paper, we propose \our{} (\underline{E}nsembling \underline{P}rioritized \underline{H}ybrid Policies), a Q-learning-based MARL-MAPF solver with communication. Our communication scheme entails an enhanced selective communication block with improvements inspired by the latest Transformer variant to enable richer information extraction. Additionally, we introduce several strategies to be used during the inference phase to further bolster performance. Firstly, we propose Q value-based priority decisions, in which the Prioritized Conflict Resolution decides the priority of agents involved in conflicts, and Advanced Escape Policy is responsible for breaking deadlocks. Secondly, we use hybrid expert guidance for agents that have no other \textit{live agents} nearby. Lastly, we utilize ensembling to sample the best solutions from multiple solvers that run in parallel to leverage the different strengths of each strategy proposed. By improving communication capability and utilizing proposed strategies to help inference, we achieve competitive performance against state-of-the-art neural MARL-MAPF solvers.

\section{Related Works}

$\textit{Search-based MAPF.}$ Traditional centralized MAPF solvers often invoke search-based algorithms, and they can be categorized by their optimality. Optimal traditional heuristics solvers include Conflict-Based Search (CBS) \cite{sharon2015conflict} and its improved version \cite{gange2019lazy,li2020new}. However, CBS relies on a constraint tree to find optimal solutions and can incur heavy computational overhead when the number of nodes in the constraint tree is high with many agents. The bounded-suboptimal solvers include CBS variants \cite{li2021eecbs, barer2014suboptimal}. These solvers can scale to environments with a larger number of agents while also guaranteeing the (sub-)optimality of generated solutions. However, their scalability is still hindered by their exponential time complexity \cite{chung2023learning}. Besides, when a new task or agent is added to the environment, such heuristics solvers have to re-plan the whole paths for all agents. 

$\textit{MARL-based MAPF.}$ Recently, various approaches have emerged to address the MAPF problem using multi-agent reinforcement learning (MARL) techniques, building upon the foundation laid by PRIMAL \cite{sartoretti2019primal}. These approaches exhibit diversity in their settings and strategies. In terms of the problem settings, some works consider scenarios with partial observation, where agents do not have complete knowledge of the environment \cite{ma2021distributed,ma2021learning,sartoretti2019primal,damani2021primal}, while others operate under full observation, assuming agents possess complete information about the environment \cite{he2021asynchronous,wang2020mrcdrl}. Furthermore, the choice of RL algorithm varies across different works. Some employ Actor-Critic approaches \cite{sartoretti2019primal,damani2021primal,wang2023scrimp}, whereas others opt for value-based Q-Learning methods \cite{ma2021distributed,ma2021learning,wang2020mrcdrl,chen2022multiagent}. In addition, certain approaches consider communication mechanisms among agents to enhance coordination \cite{ma2021distributed,ma2021learning,li2022multi,chen2022multi,wang2023scrimp}. These communication protocols allow agents to share information and collaborate effectively. Some recent MARL-based MAPF solvers also propose the use of techniques to help pathfinding during the inference phase. Gao et al. \cite{gao2023rde} propose an escape policy for structured environments that helps neural solvers escape from deadlock scenarios, which forces agents to take random actions attempting to break the deadlock. Such inference techniques can improve the performance of neural solvers under certain predefined conditions  \citep{tang2024himap}.

\section{Problem Formulation}

\subsection{MAPF Definition}
\label{subsec:mapf-definition}
The MAPF problem involves finding a set of collision-free paths for multiple agents, each with its own distinct starting location and a unique goal location, within a map containing obstacles. The map $M$ is an undirected graph $M=(V, E)$, with some vertices being inaccessible obstacles $V_o \subset V$. An obstacle collision occurs when agent $i$ reaches the location of an obstacle. Given an agent set $N$, each agent $i \in N$ has a starting vertex $v_i^0 \in V$ and a goal vertex $v_i^g \in V$. We define the observation space $O_i$ and action space $A_i$ for agent $i$, which is, at each time step $t=0,\cdots,t_{max}$, each agent $i$ located in $v^t_i \in V$ has an observation of the map $o^t_i\in O_i$, decides on a single action $a^t_{i}\in A_i$, and move to $v^{t+1}_i$. $t_{max}$ is the maximal allowed time to finish the MAPF problem. A vertex collision occurs when agent $i$ and agent $j$ reach the same vertex $v$ at the same time, step $t$; an edge conflict occurs when agent $i$ and agent $j$ traverse through the same edge $(u,v)$ in the opposite direction. A MAPF solution exists if and only if for any agent $i\in N$, there exists $0\leq t\leq t_{max}$ such that $a_i^{t}\left(\cdots a_i^{1}\left(a_i^{0}\left(v_i^0\right)\right)\right)=v_i^{t+1}=v_i^g$, where $a^0_i(v_i^0)=v^1_i, ..., a^{t}_i(v_i^{t})=v_i^{t+1}=v_i^g$; and for any agent $i\in N$, there are no collisions.

\begin{figure*}
    \centering
        \vspace{3mm}

    \begin{tikzpicture}
        \node[anchor=south west,inner sep=0] (image) at (0,0) {\includegraphics[width=0.97\linewidth]{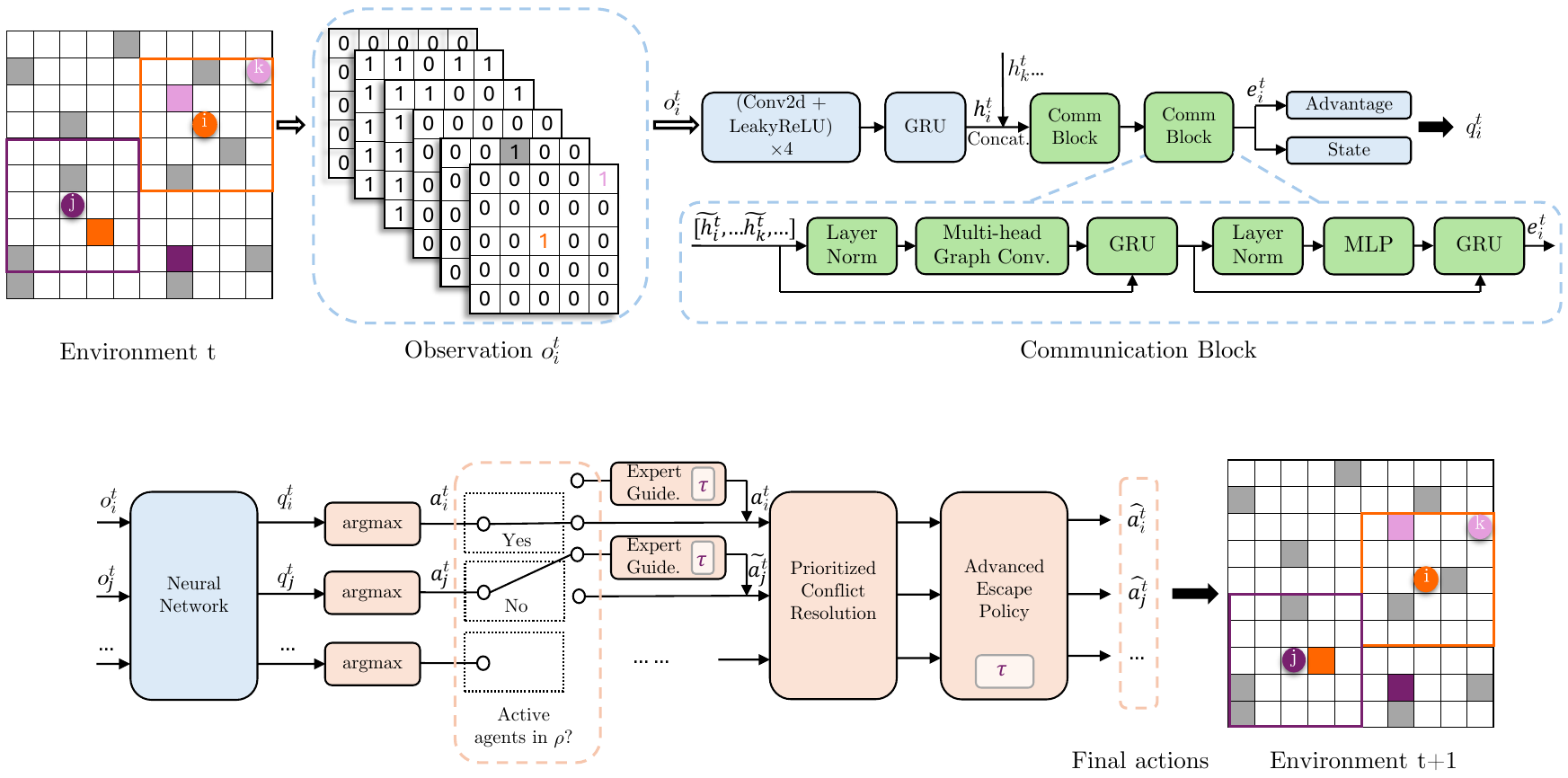}};
        \begin{scope}[x={(image.south east)},y={(image.north west)}]
            \node[anchor=north, fill=white, draw=black, text width=5cm, align=center, rounded corners=5pt] at (0.5,1.00) {\small Neural Network (\cref{subsec:qlearningmapf})};
            \node[anchor=north, fill=white, draw=black, text width=5cm, align=center, rounded corners=5pt] at (0.5,0.5) {\small Inference Strategies (\cref{subsec:collaboration})};
        \end{scope}
    \end{tikzpicture}
    \caption{Overview of a single inference step of \our{}. The upper part is the neural network structure of \our{}; the lower part is the illustration of how observations of all agents are transformed into actions by first feeding into the neural network, and then going through the proposed inference strategies. The $\rho$ and $\tau$ in the lower part are the hyperparameters defined in \cref{subsec:collaboration}.}
    \label{fig:eph}
\end{figure*}

\subsection{Environment Settings}
\label{subsec:environment}
Following the conventions of MAPF, we use a 2D grid world to represent the environment. Each cell in the $n \times n$ grid world can be either an empty cell or an obstacle that stops agents from passing\footnote{We consider value $1$ on map $M$ as an obstacle and $0$ as available.}. Each agent occupies an empty cell, and the task for each agent is moving from its starting cell to the goal cell without collisions. At the beginning of each episode, $m$ starting cells and $m$ goal cells are randomly selected among the empty cells for $m$ agents, and we make sure that there is no overlap among $2m$ selected cells. At each timestep, the agents choose to move either up, down, left, right, or stay still. We consider four types of conflicts: \textit{edge conflict} \cite{stern2019multi}, \textit{vertex conflict} \cite{stern2019multi}, conflict with static obstacles, and out-of-bound conflict. These four types of conflicts are not allowed during pathfinding. 

We use a partially observable environment. At each time step, each agent has a limited Field of View (FOV) that consists of six channels: four channels are the heuristic channels described in \cite{ma2021distributed}, the remaining two channels are used to represent the locations of all agents within the FOV, and the location of obstacles within the FOV, respectively.

\section{\our{}: Ensembling Prioritized Hybrid Policies}
\label{sec:eph}
 We propose \our{} and describe it in two sections, each with several subsections. \cref{subsec:qlearningmapf} describes how we formulate MAPF problem as a Q-Learning based MARL problem with communication and how we train such a model. \cref{subsec:collaboration} gives an introduction to the advanced inference strategies that we propose to improve performance during execution. \cref{fig:eph} shows the overall architecture of \our{}.

 \subsection{Q-Learning based MARL with Communication for MAPF}
\label{subsec:qlearningmapf}

 In this section, we first introduce our model with the new communication block, and then we introduce the model training via Double Dueling Deep Q Networks (D3QN).
\subsubsection{Graph Convolution based Communication}
\label{subsubsec:method-communication}

We show our communication architecture in \cref{fig:eph}. Our communication structure is adapted from selective communication block in DCC \cite{ma2021learning}. Compared with ordinary communication, selective communication can effectively reduce communication time compared with non-selective counterparts, it can also selectively gather more informative and important hidden information from other agents within FOV, which is shown to be beneficial for pathfinding \cite{ma2021learning}. However, in the previous work, the selective graph convolution communication block did not include position-wise processing or normalization, and it is shown that including them helps stabilize training and gives better scalability \cite{parisotto2020stabilizing}. Inspired by $\textit{GTrXL}$-styled transformer in \cite{parisotto2020stabilizing}, we re-design the communication block to enable better coordination among agents by including an extra multi-layer perception layer, GRU block, and layer normalization, which is shown to be beneficial in many works across different fields \cite{vaswani2017attention,wang2023scrimp,parisotto2020stabilizing}. The generation method of communication mask for selective communication, as well as the request-reply communication scheme that consists of two cascaded communication blocks, are kept the same as shown in DCC to reduce communication overhead and enrich the information gathered.

\subsubsection{Training of Q-learning based MARL with Communication}
\label{subsubsec:method-training}
We train the model with D3QN (Double Dueling Deep Q Network), which incorporates Deep Q-Learning Network (DQN) with both the Dueling DQN architecture \citep{wang2016dueling}, separating parameters for advantage and value estimation, and Double DQN \citep{van2016deep}, which uses a target network to estimate the value to ameliorate the effects of Q value overestimation \citep{hasselt2010double}. The Q value for agent $i$ is obtained via:
\begin{equation}
Q_{s,a}^{i} = Val_s\left(e_{i}^{t}\right) +  Adv\left(e_{i}^{t}\right)_a - \frac{1}{\left|\mathcal{A}\right|} \sum_{a'} Adv\left(e_{i}^{t}\right)_{a'}
\end{equation}
where $Val(\cdot)$ and $Adv(\cdot)$ represents state advantage and action advantage function, respectively. $e_{i}^{t}$ is the final output of communication for agent $i$ containing information from other agents whom $i$ chooses to communicate with. $\mathcal{A}$ is the action space. The loss for training \our{} is expressed as:
\begin{equation}
\mathcal{L}(\theta) = \text{MSE} \left( R_t^i - Q_{s_t,a_t}^i (\theta) \right) 
\end{equation}
where MSE is the mean square error. We have $R_t^i = r_t^i + \gamma r_{t+1}^i + \ldots + \gamma^n Q_{s_{t+n}, a_{t+n}}^i(\bar{\theta})$, where $r_t^i$ is the reward received by agent $i$ at time $t$, and $\bar{\theta}$ is the parameter for target network which is updated to align with online parameter $\theta$ at a predefined interval. The reward structure for D3QN is taken from DHC \cite{ma2021distributed} as shown in the \cref{table:reward}. Negative rewards are given to agents who don't reach the goal to expedite goal reaching in the shortest distance. 

\input{tables/rewards.tex}

In MAPF, the roles of each agent are identical: every agent has a unique start location and a goal location, and each of them is expected to move in a collision-free manner and reach the goal eventually. Given this observation, instead of training multiple policies for multiple agents, it's more natural to adopt parameter sharing and train a single policy from a single agent's perspective. Although we adopt parameter sharing, it should be emphasized that the trained policy can be used for multi-agent pathfinding. This is because each agent perceives unique information from its own partial observations and inter-agent communication outcome and, consequently, takes different actions.

\subsection{Advanced Inferences Strategies for \our{}}
\label{subsec:collaboration}

We propose three inference techniques to further bolster pathfinding performance during inference. In \cref{subsubsec:method-experguidance}, we hybridize the policy with single-agent optimal $A^*$ paths to give guidance to agents that do not have any other $\textit{live agents}$ nearby. In \cref{subsubsec:method-priorities}, we introduce two Q value-based priority decision-making strategies, namely \textit{Prioritized Conflict Resolution} for efficient conflict resolution and \textit{Advanced Escape Policy} for priority-based avoidance of deadlocks. Finally, \cref{subsubsec:method-ensembling} introduces the ensembling, which runs multiple policies in parallel and samples the best possible solutions in the solution space. 
\subsubsection{Hybrid Expert Guidance}
\label{subsubsec:method-experguidance}

When agent communication is sparse, such as in scenarios where no other agents are located within the agent's field of view, it is beneficial to incorporate low-cost single-agent expert path to guide decision-making, thus hybridizing (EP\underline{H} stands for \underline{H}ybrid). We hereby define a $\textit{live agent}$ as an agent that is currently off its goal, i.e., $v_i^t \neq v_i^g$. We propose leveraging the optimal $A^*$ path as expert guidance during inference for an agent that has no other $\textit{live agents}$ within a square centered on the agent itself; the length of the square is $2\rho+1$ where $\rho$ is the visibility radius for $\textit{live agents}$. We adapt the efficient $A^*$, a well-established low-cost algorithm for efficient single-agent pathfinding, to multi-agent settings as follows:
\begin{equation}
     a_i^{*t}, a_i^{*t+1}, \dots, a_i^{*T} =  A^*_\tau (v_i^t, v_i^g, M, V^t, V^g)
\end{equation}
where $a_i^{*t}, a_i^{*t+1}, \dots, a_i^{*T}$ is the sequence of actions that leads agent $i$ to the goal and $\tau \in \{0, 1, 2\}$ is the $A^*$ type. $V^t=\left \{ v_1^t,...,v_i^t,...,v_m^t \right \}$ and $V^g=\left \{ v_1^g,...,v_i^g,...,v_m^g \right \}$ is the current position set and goal set, respectively. We formulate each type $\tau$ as follows:

\begin{enumerate}
    \item $A^*_0$: is the classic $A^*$ that takes the current map $M$ for avoiding obstacles ensuring finding the optimal single-agent path. 
    \item $A^*_1$: treats \textit{all agents as obstacles}, except the current agent $i$; i.e., $M(v_j^t) \gets 1 ~ \forall j, ~and j \neq i$. All other agents are considered temporary obstacles within the $A^*$ calculation. This ensures collision avoidance.
    \item $A^*_2$: this version considers \textit{only inactive agents as obstacles}, i.e., $M(v_j^t) \gets 1 ~ \forall j ~s.t.~ v_j^t = v_j^g$. This version tries to avoid agents already at goal only. This can generally obtain better paths compared to $A^*_1$, since agents that are still moving do not interfere with optimal path calculations.
\end{enumerate}

\begin{figure}%
    \centering
    \vspace{3mm}
    \includegraphics[width=0.5\linewidth]{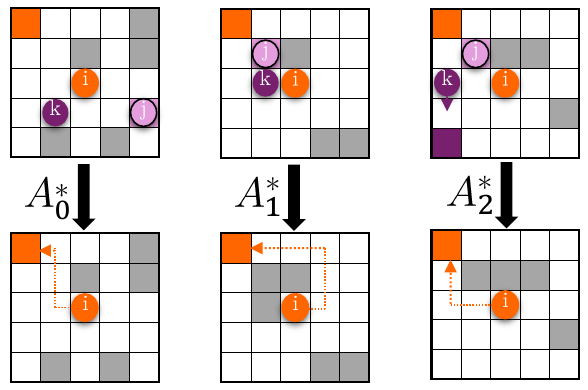}
    \caption{Types of $A^*_\tau$ we consider. $\tau \in \{0,1,2\}$. Changing the map representation improves the single-agent path based on each situation.}
    \label{fig:astar}
\vspace{-1em}
\end{figure}

The illustration of three types of $A^*$ is shown in \cref{fig:astar}. We find that using different $A^*$ types can be beneficial for pathfinding in different scenarios.

\subsubsection{Value-based Priority Decisions} 
\label{subsubsec:method-priorities}

We further introduce two techniques for resolving conflict situations and deadlock scenarios based on priorities (E\underline{P}H stands for \underline{P}rioritized)  provided by the Q values obtained by the trained neural model. This is inspired by priority-based planning methods \citep{latombe2012robot, silver2005cooperative, okumura2022priority} and, in particular, Priority-Based Search (PBS) \cite{ma2019searching}, which first decides the priority for each agent and then plans individual paths for each agent in the order of priority. In PBS, the agent with lower priority is not allowed to collide with the agent with higher priority. In our Q value-based priority decisions, an agent with a lower Q value will be assigned a lower priority and will have to give way to another agent with a higher priority. Intuitively, higher Q-values correspond to higher expected rewards.

\begin{figure}%
    \centering
    \includegraphics[width=0.5\linewidth]{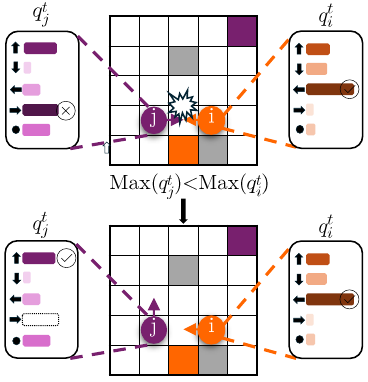}
    \caption{Prioritized Conflict Resolution. Agents with higher Q values are prioritized when a conflict happens, which leads to shorter paths.}
    \label{fig:priority}
\vspace{-1em}
\end{figure}

\paragraph{Prioritized Conflict Resolution}
During pathfinding, agents may still take invalid actions, which leads to conflicts. Compared with previous works \cite{ma2021distributed,ma2021learning} that recursively recover the states of agents that are involved in conflicts, we use several strategies to reduce collisions. When we first sample actions from the Q value matrix generated by our network, we filter out actions that make agents collide with static obstacles by masking the corresponding value in the Q value matrix. After we get static-obstacle-collision-free action for each agent, we let all agents execute their actions. If collisions between agents happen, we decide the priority of agents involved in agent collision by giving the highest priority to the agent with the highest Q value, and re-choosing actions for other agents with smaller Q values with their previous actions masked. The illustration of Prioritized Conflict Resolution is shown in \cref{fig:priority}. Since Q value represents the goodness and preference of an agent choosing the corresponding action, it is natural to decide the priority of agents based on Q value.

\paragraph{Advanced Escape Policy} Gao et al. \cite{gao2023rde} first proposed an escape policy to avoid deadlock situations, a common case in structured environments. A deadlock can occur when an agent is off its goal and oscillates back and forth between two adjacent cells (\cref{alg:aep}, lines 9-10). However, in the original escape policy in \cite{gao2023rde}, agents choose to break deadlocks just by taking random actions when deadlocks are detected, which may be suboptimal and lead to unnecessary randomness\footnote{With time $\to \infty$, a series of random actions should also guarantee a solution at the expense of solving time.}. In our Advanced Escape Policy, we use a two-stage approach to break the deadlocks. Firstly, we assign a priority to each agent by sorting the Q values in descending order. Then, for each agent in such order, we utilize $A^*_\tau$ mentioned in \cref{subsubsec:method-experguidance} to obtain the action. If $A^*_\tau$ fails to find a path because of other agents involved in the deadlock, we choose the action from the highest \textit{valid} Q value, where \textit{invalid} Q values (corresponding to actions that lead to a collision) are masked, i.e. set to $-\infty$. We also update a copy of the map such that the current next action is masked, ensuring there will be no conflict. \cref{alg:aep} shows our proposed Advanced Escape Policy.

\input{algo_aep}

\input{fig/main_results}

\subsubsection{Ensembling}
\label{subsubsec:method-ensembling}

In the inference phase, \underline{E}PH (where \underline{E} stands for the ensembling) employs an ensemble technique that operates by sampling best solutions from solvers with different settings that run in parallel, without engaging in adaptive learning. Each of the solvers has its own $A^*$ type, either $A^*_0$, $A^*_1$ or $A^*_2$, with different $\rho$. Some of the solvers have the value-based priority decision mechanism enabled, while others don't. This approach is designed to leverage the complementary strengths of each strategy proposed to navigate complex multi-agent pathfinding. We can divide the ensembling technique into two stages:

\begin{enumerate}

        \item \textit{Parallel Solver Execution:} For every episode, \our{}  executes multiple solvers with different settings in parallel, each of them has different techniques enabled with different parameters. We treat each solution from one solver in the parallel execution as one independent pathfinding solution. Parallel execution allows for a comprehensive exploration of potential solutions, maximizing the likelihood of identifying an optimal path.
    
    \item \textit{Best Solution Sampling:} Upon completion of all parallel solvers, \our{} evaluates each independent solution from one solver based on its \textit{makespan} \cite{stern2019multi}, i.e., time steps required to finish the single test episode. For each episode, we take as a final solution the one with the lowest \textit{makespan} among all parallel solvers.
    
\end{enumerate}

This ensemble technique enhances \our{}'s capability to solve MAPF problems by providing a robust framework for selecting the most effective pathfinding solver from a set of pre-defined options. By systematically sampling the best solutions from multiple solvers running in parallel, \our{} efficiently navigates the complexities of multi-agent environments, ensuring high performance and reliability in the inference phase.

\section{Experiments}

\input{tables/structure_maps}

\subsection{Training and Testing Settings}
We use D3QN with a prioritized experience replay buffer \citep{schaul2015prioritized} for training \our{}. Curriculum learning \cite{bengio2009curriculum}, which is shown to be effective in previous works \cite{ma2021distributed,lin2023sacha}, is also adopted. The curriculum learning starts from one agent in $10 \times 10$ map, and ends on $40 \times 40$ map with 16 agents. Every time the success rate of the model under training reaches 0.9 for the current training environment, we increase the difficulty of the map. Thus, experience from more complex environments will be added to the buffer, and the model learns from more and more complex scenarios gradually. The obstacle density of maps in training is sampled from a triangular distribution between 0 and 0.5 with a peak of 0.33. The FOV size in our work is $9 \times 9$ to align with previous works \cite{ma2021distributed,ma2021learning}. The maximum episode length for training episodes is 256. We save model checkpoints every 1000 training steps and take the best one as the final model based on performance on a random-map validation set. It takes 20 hours to perform 150k training steps on a server with a single \textsc{NVIDIA RTX A6000} and \textsc{AMD EPYC 7542 (128) @2.9GHz}. We also retrain DCC with the same setting for fair comparison. 

We evaluate \our{} on random maps and structured environments. For random maps, we test on $40 \times 40$ and $80 \times 80$ maps with 0.3 obstacle density; the maximum allowed time step for them is 256 and 386, respectively. The set of parameters $\tau$ and $\rho$ used by ensemble for random maps are selected from a Cartesian product of the spaces $\{0,1,2\} \times \{2,3,4,5\}$. For structured environments, we use the benchmark from \cite{stern2019multi} for the test. Specifically, we choose the Dragon Age Origins map ${\texttt{den312d}}$ ($65 \times 81$) and ${\texttt{warehouse}}$ map ($161 \times 63$) from the benchmark for the test to showcase \our{}'s performance in complex structured environments. The maximum allowed time step for ${\texttt{den312d}}$ and ${\texttt{warehouse}}$ is 256 and 512, respectively. The set of parameters $\tau$ and $\rho$ used by ensemble for structured maps are selected from a Cartesian product of the spaces $\{0,1,2\} \times \{3,4\}$. For random maps, we report the success rate and episode length averaged across 100 test instances for each case. For structured maps, we average across 300 to align with the setting in one of our baseline SACHA \cite{lin2023sacha}. The success rate is the percentage of test instances fully solved, and the episode length is the \textit{makespan} \cite{stern2019multi} of an instance.

\subsection{Results on Random Maps}
We choose DHC, DCC, and current state-of-the-art MARL-MAPF solver SCRIMP \cite{wang2023scrimp} as our baselines. DHC and DCC both utilize local communication to gather information from other agents nearby to give the model more information and understanding beyond the agent's own FOV, and both use Q-learning. While DHC chooses to communicate with all other agents nearby, DCC selectively communicates with them. SCRIMP uses Transformer-based global communication to even gather more information from all agents in the environment to enrich the information one agent can receive. Imitation learning is also utilized to help SCRIMP learn from expert demonstrations from multi-agent solvers during training alongside PPO \citep{schulman2017proximal}. 

\cref{fig:episodelength-successrate} (left) shows the success rate of \our{} compared with baselines. On $40 \times 40$ map, \our{} beats all the baselines in terms of success rate, reaching $100\%$ success rate for all the cases; on the $80 \times 80$ map, \our{} outperforms both DHC and DCC in all cases, while falling behind SCRIMP by $1\%$ on 32 and 64 agents scenarios. \cref{fig:episodelength-successrate} (right) illustrates the average episode length, which indicates the solution quality, of all four models. \our{} outperforms both DHC and DCC in terms of average episode length in all cases. When compared against SCRIMP, \our{} can outperform SCRIMP on larger $80 \times 80$ map. On $40 \times 40$ map, \our{} can achieve a shorter average episode length than SCRIMP except in the 64-agent case. It should be noted that SCRIMP uses global communication, which could give the model more information than we could. However, global communication is burdensome and inefficient in real life; also, global communication may be limited by geographical constraints. By contrast, \our{}, which achieves better performance than baselines in most cases, only uses improved local selective communication. \our{} equipped with local selective communication is information-efficient, less burdensome, and practical in real-life scenarios.

\subsection{Results on Structured Maps}
We additionally test on structured maps that were never seen during training to showcase the performance of baselines and \our{} in terms of scalability and generalization to unseen scenarios. For structured maps, apart from the abovementioned baselines, we further include three additional heuristic solvers: CBS \cite{sharon2015conflict}, wPBS \cite{li2021lifelong} and ODrM$^*$ \cite{ferner2013odrm}; as well as two neural solvers: PRIMAL \cite{sartoretti2019primal} and the recent SACHA \cite{lin2023sacha} as baselines. SACHA is a neural MAPF solver trained with Soft Actor-Critic and employs optional global communication blocks. We choose to compare with SACHA equipped with global communication block for fair comparison. For the heuristics solvers, the runtime limit is 120 seconds for CBS and wPBS, and 20 seconds for ODrM$^*$, same as the settings used in SACHA. \our{} takes a slightly higher runtime compared to DCC. At the same time, conflict checks and running A$^*$ have a negligible effect on runtime\footnote{We note that we did not parallelize ensembling in the implementation; parallelizing ensembling would significantly speed up the algorithm. We leave this implementation for future work.}.

\cref{tab:average_step} shows the performance of different solvers in two structured environments. In ${\texttt{den312d}}$ map, \our{} beats all neural baselines in terms of success rate. For the average episode length, we outperform all neural baselines except SACHA in ${\texttt{den312d}}$ map with 8 to 32 agents. It is noticeable that though SACHA achieves shorter episode length in these cases, it fails to achieve a higher success rate when the number of agents increases. By contrast, \our{} has better scalability and achieves $100\%$ success rate in ${\texttt{den312d}}$ map with 64 agents. In the ${\texttt{warehouse}}$ map, which is a common setup in real-world scenarios, \our{} excels all neural baselines in all cases in terms of both metrics, showcasing the practicality of our method in real-world applications. It is noticeable that heuristic solvers generally offer better solutions in cases with few agents. However, the scalability issue, which is common for heuristics solvers, prevents them from reaching better results when the agent number increases.

\subsection{Ablation Studies}
\input{tables/ablation_table}

In \cref{table:ablation}, we showcase the effectiveness of the Q value-based priority decision-making strategy, hybrid expert guidance, and ensembling outlined in \cref{subsec:collaboration}. \textit{Base} is the model that only has the proposed improved communication block. We use $A^*_2$ in the relative parts of value-based priority decision-making strategy and hybrid expert guidance. In \texttt{den312d}, it can be seen that all three inference strategies help achieve better performance compared with \textit{Base}. In \texttt{warehouse}, the same trend is preserved. However, we observe that different strategies have different improvements in different scenarios. For example, hybrid guidance does not offer much improvement in maps that are not highly congested and structured, such as \texttt{den312d}; but in highly structured environments like \texttt{warehouse}, it offers significant improvements. This signals that ensembling to take advantage of different solvers would be the best choice to give the model the best performance. In addition, it is also noticeable that the \textit{Base} outperforms DCC in both structured maps, illustrating the effectiveness of our improved communication scheme.

We also conduct an ablation study regarding the impact of $A^*$ types of the hybrid expert guidance, i.e., the $\tau$, on \our{}'s performance. We observe that in \texttt{den312d} map with $64$ agents, different $\tau$ only leads to $4\%$ of difference in success rate. However in \texttt{warehouse} map with $64$ agents, the difference jumps to $79\%$, and the model with hybrid expert guidance with $A^*_0$ only achieves $18\%$ success rate. This signifies again that different strategies have their own applicability and the effectiveness of ensembling.

\section{Conclusions}

In this paper, we proposed \our{}, a novel learning approach for Multi-Agent Path Finding (MAPF). Our approach employs an enhanced selective communication scheme and, at inference time, efficiently samples multiple solutions by ensembling solvers with configurations of neural value-based priorities for resolving conflicts, advanced escape policy for deadlock avoidance, and hybrid expert guidance. \our{} demonstrated competitive performance against state-of-the-art neural MAPF baselines in complex environments. 

We finally define some future works that may arise from \our{}. Firstly, training with other RL algorithms such as on-policy algorithms \citep{williams1992simple, schulman2017proximal, berto2024rl4co}, where priorities may be assessed via values from the critic network, could further boost the performance.
Importantly, better hybridization techniques with existing inexpensive low-level solvers may help in particular by avoiding deadlock situations arising in highly-structured environments \cite{damani2021primal}. 
Ultimately, we believe hybridizing learned neural solvers with their classical counterparts, as has been suggested in other combinatorial domains for both obtaining better solutions \citep{hottung2019neural,kool2022euro, ye2023glop, ye2024deepaco} and generating better heuristics \citep{liu2024example, ye2024reevo}, is a promising research avenue for scalable and generalizable MAPF.

\section*{ACKNOWLEDGMENT}
We thank Qiushi Lin for providing us with help for the performance results of baselines in structured maps.

\bibliographystyle{IEEEtran}
\bibliography{bibliography}

\clearpage

\end{document}

%% file: tables/rewards.tex
\begin{table}[ht]
\centering
\caption{Reward Structure}
\begin{tabular}{@{}l | c@{}}
\toprule
\textbf{Actions} & \textbf{Rewards} \\ 
\midrule
\ding{224} Move and Stay off Goal & -0.075 \\ %
\ding{52} Stay on Goal           & 0 \\ %
\ding{54} Collision              & -0.5 \\ %
\ding{72} Reach Goal             & 3 \\ %
\bottomrule
\end{tabular}
\label{table:reward}
\vspace{-1em} %
\end{table}

%% file: algo_aep.tex
\begin{algorithm}[ht]
\small
\caption{Advanced Escape Policy}
\label{alg:aep}
\KwIn{Goals set $V^g$; agents' current positions set $V^t$; position set for each agent for the past five time steps $ V^p_i=\left \{ v^{t}_i,v^{t-1}_i,v^{t-2}_i,v^{t-3}_i,v^{t-4}_i \right \}; $ Q value matrix $Q=\left ( q^t_1 \cdots q^t_i \cdots q^t_m \right ), ~ \forall i = 1, \dots, m$; initial map $M$ where $1$ is an obstacle and $0$ is available; $A^*$ type $\tau$.}
\KwOut{Actions $A=\left \{ a^t_1,...,a^t_i,...,a^t_m \right \}$
}

\SetKwFunction{FUpdateQ}{MaskQValues}
\SetKwProg{Fn}{Function}{:}{}
\Fn{\FUpdateQ{$q_i^t, M, v_i^t$}}{
    \tcp{mask Q values of obstacles}
    \For{$a$ \text{\textbf{in}} $\mathcal{A}$}{
        \uIf{$M[a(v_i^t)] = 1$}{
            $q_i^t(a) \gets -\infty$\;
        }
    }
    \KwRet $q_i^t$\;
}
    
\tcp{Init actions based on max Q values}
${A}^{\text{init}} = \left \{ a^t_1,...,a^t_i,...,a^t_m \right \} \gets \text{arg}\max{Q}$ \;
$Idx \gets \text{argsort}(Q_{{A}^{\text{init}}},\text{descending=True)}$ \tcp*{sort indexes by Q values (first is highest)}

\For{$i$ in $Idx$}{

	\tcp{check if active and has deadlock}
	\If{$v^t_i \neq v^g_i$}{
		\If{$v^{t-1}_i = v^{t-3}_i \And v^{t-2}_i = v^{t-4}_i$}
  { 
           \tcp{if $A^*_\tau$ path found, use $A^*_\tau$}
        
            \If{$\exists ~A^*_\tau(v_i^t, v_i^g, M, V^t, V^g)$}{
            	$a^{t}_i \gets ~A_\tau^*(v_i^t, v_i^g, M, V^t, V^g)[0]$
            }
            \Else{
            	$q_i^t \gets$ \FUpdateQ{$q_i^t, M, v_i^t$}\;
            	$a^{t}_i \gets \text{arg}\max{q_i^t}$
            }
		}
	}
    
    $M[a_i^t(v_i^t)] \gets 1$ \tcp{set next pos. as obstacle}
}

\end{algorithm}

%% file: fig/main_results.tex
\begin{figure*}[t] %
    \vspace{3mm}
    \centering
    \begin{subfigure}{0.48\linewidth}
        \includegraphics[width=\linewidth]{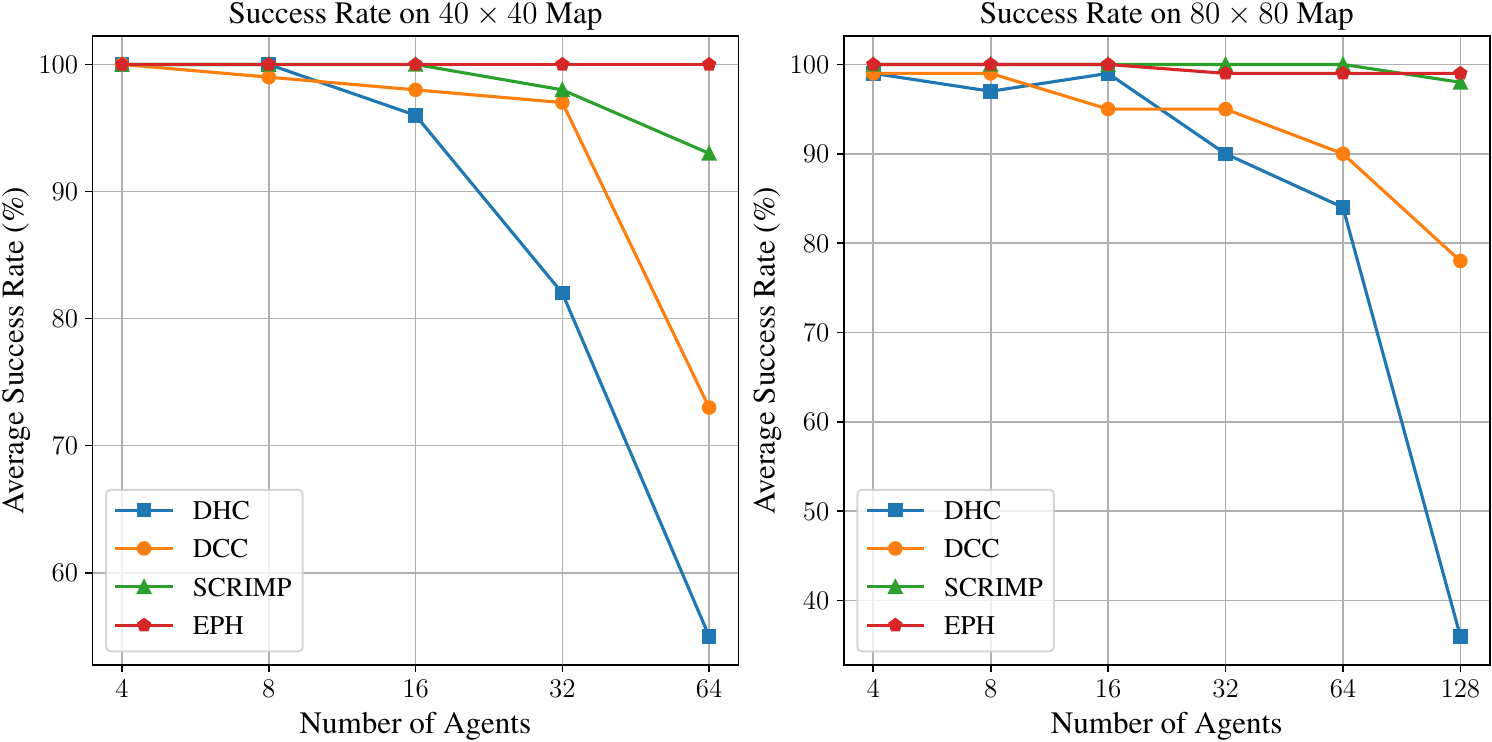}
    \end{subfigure}
    \hfill %
    \begin{subfigure}{0.48\linewidth}
        \includegraphics[width=\linewidth]{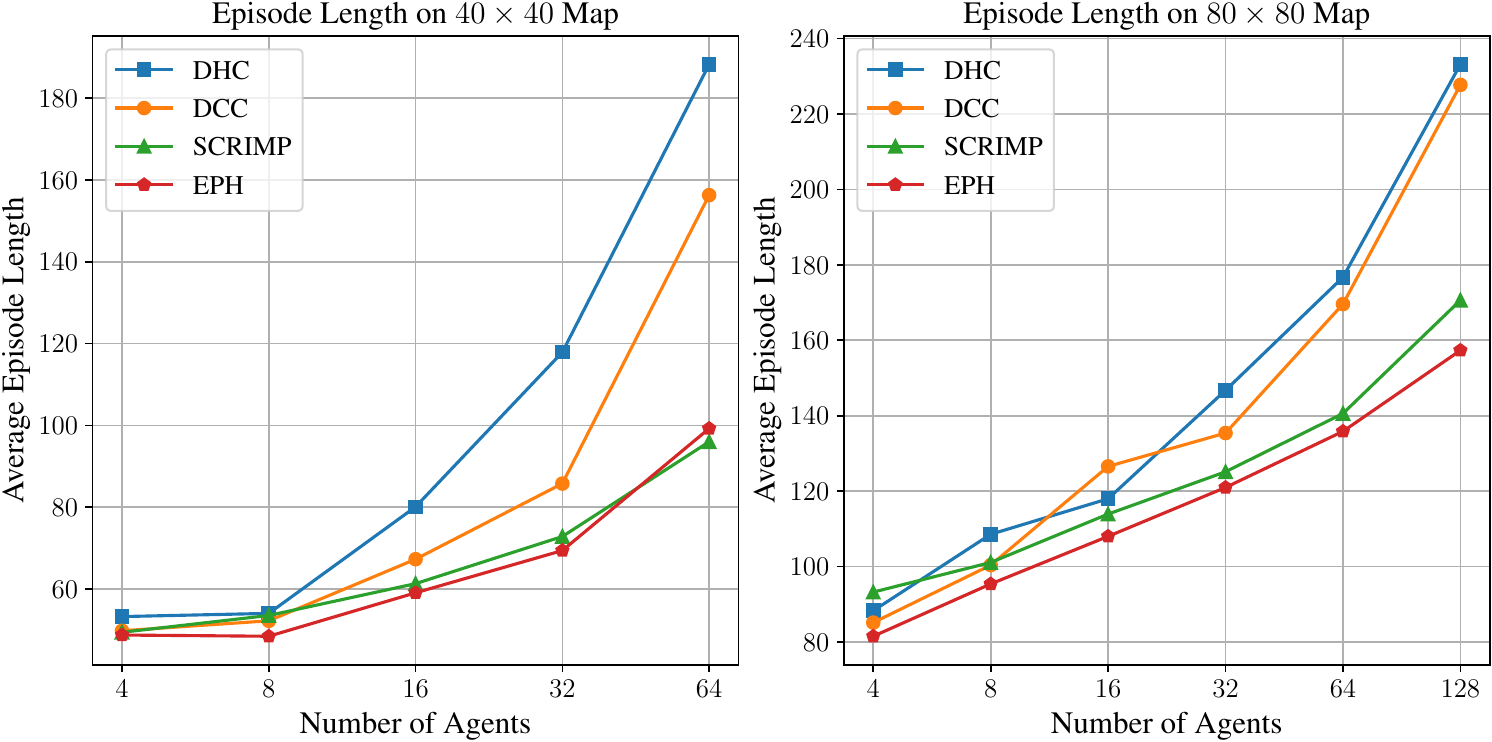}
    \end{subfigure}
    \caption{Comparative Analysis of Success Rate (left) and Average Episode Length (right) on random maps.}
        \label{fig:episodelength-successrate}
\end{figure*}

%% file: tables/structure_maps.tex
\begin{table*}[ht]
\centering
\caption{Solution quality for different MAPF solvers. We report episode length (EL, lower is better $\downarrow$) and success rate (SR, higher is better $\uparrow$).}
\label{tab:average_step}
\scriptsize
\setlength{\tabcolsep}{4pt}
\resizebox{\textwidth}{!}{
\begin{tabular}{c|c|*{3}{| c c} |   *{6}{| c c}}
\toprule

\multicolumn{2}{c}{} & 
\multicolumn{6}{c}{Heuristics Solvers} & \multicolumn{12}{c}{Neural Solvers}
\\
\cmidrule(lr){3-8}  \cmidrule(lr){9-20}

  \multicolumn{1}{c}{} & \multicolumn{1}{c}{}  & \multicolumn{2}{c}{CBS} & \multicolumn{2}{c}{ODrM*} & \multicolumn{2}{c}{wPBS} & \multicolumn{2}{c}{PRIMAL} & \multicolumn{2}{c}{DHC} & \multicolumn{2}{c}{DCC} & \multicolumn{2}{c}{SACHA} & \multicolumn{2}{c}{SCRIMP} & \multicolumn{2}{c}{EPH} \\
\cmidrule(lr){3-4} \cmidrule(lr){5-6} \cmidrule(lr){7-8} \cmidrule(lr){9-10} \cmidrule(lr){11-12} \cmidrule(lr){13-14} \cmidrule(lr){15-16} \cmidrule(lr){17-18} \cmidrule(lr){19-20}
 Map & $m$  & \begin{tabular}[c]{@{}c@{}}EL\end{tabular} & \begin{tabular}[c]{@{}c@{}}SR\end{tabular} & \begin{tabular}[c]{@{}c@{}}EL\end{tabular} & \begin{tabular}[c]{@{}c@{}}SR\end{tabular} & \begin{tabular}[c]{@{}c@{}}EL\end{tabular} & \begin{tabular}[c]{@{}c@{}}SR\end{tabular} & \begin{tabular}[c]{@{}c@{}}EL\end{tabular} & \begin{tabular}[c]{@{}c@{}}SR\end{tabular} & \begin{tabular}[c]{@{}c@{}}EL\end{tabular} & \begin{tabular}[c]{@{}c@{}}SR\end{tabular} & \begin{tabular}[c]{@{}c@{}}EL\end{tabular} & \begin{tabular}[c]{@{}c@{}}SR\end{tabular} & \begin{tabular}[c]{@{}c@{}}EL\end{tabular} & \begin{tabular}[c]{@{}c@{}}SR\end{tabular} & \begin{tabular}[c]{@{}c@{}}EL\end{tabular} & \begin{tabular}[c]{@{}c@{}}SR\end{tabular} & \begin{tabular}[c]{@{}c@{}}EL\end{tabular} & \begin{tabular}[c]{@{}c@{}}SR\end{tabular}\\
\midrule
\multirow{5}{*}{\includegraphics[width=14mm]{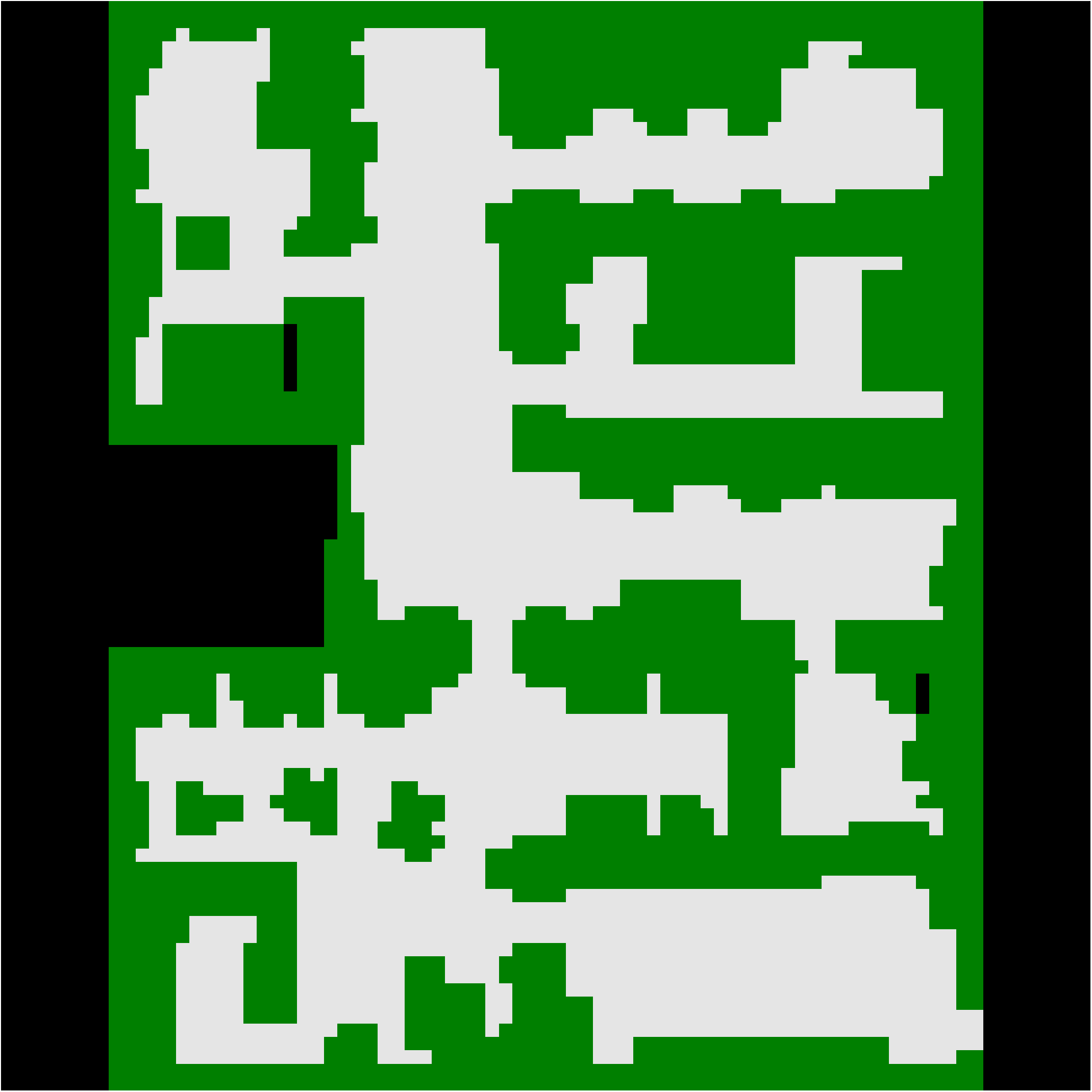}~\rotatebox[origin=t]{90}{\hspace{11mm}\texttt{den312d}}}  & 4 & 51.74 & 100\% & 51.76 & 100\% & 69.32 & 92\% & 196.54 & 40\% & 86.56 & \textbf{100\%} & 82.99 & \textbf{100\%} & 81.43 & \textbf{100\%} & 82.34 & \textbf{100\%} & \textbf{79.05} & \textbf{100\%} \\
 & 8 & 55.50 & 100\% & 78.74 & 88\% & 116.32 & 70\% & 245.02 & 8\% & 100.70 & \textbf{100\%} & 97.95 & 99\% & \textbf{89.73} & \textbf{100\%} & 99.58 & \textbf{100\%} & 91.42 & \textbf{100\%} \\
 & 16 & 118.97 & 68\% & 186.44 & 34\% & 208.28 & 24\% & \sout{256.00} & 0\% & 109.24 & \textbf{100\%} & 108.29 & 97\% & \textbf{96.74} & \textbf{100\%} & 105.78 & \textbf{100\%} & 104.87 & \textbf{100\%} \\
 & 32 & 251.86 & 2\% & \sout{256.00} & 0\% & 248.06 & 4\% & \sout{256.00} & 0\% & 124.38 & 98\% & 119.15 & 97\% & \textbf{104.30} & 98\% & 115.39 & \textbf{100\%} & 110.78 & \textbf{100\%} \\
 & 64 & \sout{256.00} & 0\% & \sout{256.00} & 0\% & \sout{256.00} & 0\% & \sout{256.00} & 0\% & 153.17 & 93\% & 145.21 & 93\% & 142.97 & 94\% & 131.59 & \textbf{100\%} & \textbf{121.66} & \textbf{100\%} \\
\midrule
\multirow{5}{*}{\includegraphics[width=14mm]{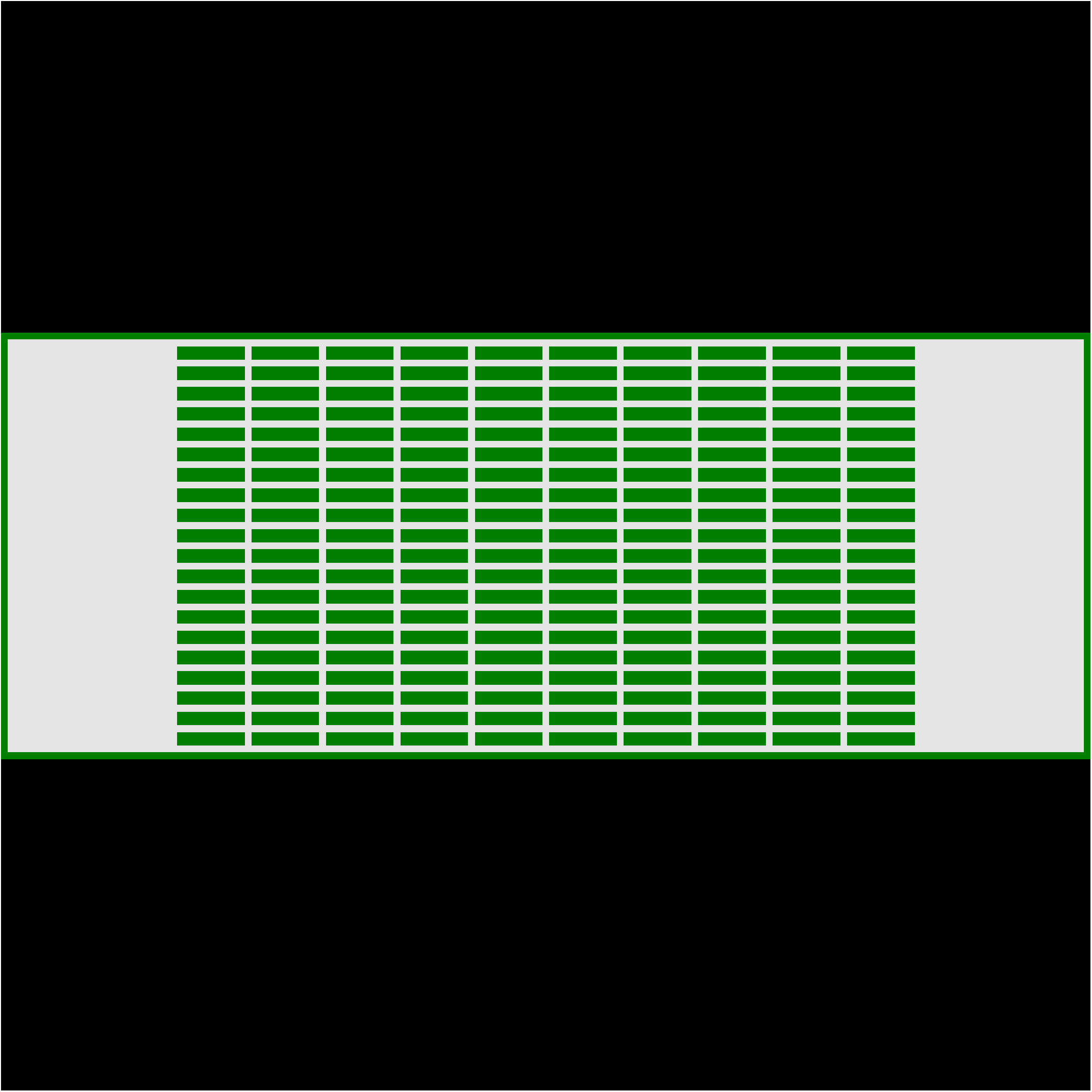}~\rotatebox[origin=t]{90}{\hspace{11mm}\texttt{warehouse}}} & 4 & 77.79 & 100\% & 77.79 & 100\% & 104.41 & 94\% & 355.80 & 42\% & 146.12 & 99\% & 135.89 & 99\% & 134.59 & 99\% & 197.79 & 87\% & \textbf{134.56} & \textbf{100\%} \\
 & 8 & 83.48 & 100\%& 100.37 & 96\% & 170.46 & 80\% & 451.82 & 18\% & 198.82 & 91\% & 169.50 & 96\% & 166.72 & 93\% & 304.22 & 66\% & \textbf{151.94} & \textbf{100\%} \\
 & 16 & 81.64 & 100\% & 133.59 & 88\% & 340.18 & 40\% & 492.04 & 8\% & 281.37 & 74\% & 208.72 & 90\% & 198.72 & 76\% & 366.98 & 48\% & \textbf{164.05} & \textbf{100\%} \\
 & 32 & 262.15 & 58\% & 417.22 & 22\% & \sout{512.00} & 0\% & 505.58 & 4\% & 432.28 & 28\% & 335.81 & 58\% & 354.33 & 48\% & 451.40 & 21\% & \textbf{176.35} & \textbf{100\%} \\
 & 64 & 494.93 & 4\% & \sout{512.00} & 0\% & \sout{512.00} & 0\% & \sout{512.00} & 0\% & \sout{512.00} & 1\% & 473.92 & 14\% & 437.29 & 28\% & 504.26 & 4\% & \textbf{189.58} & \textbf{100\%} \\
\bottomrule
\end{tabular}}
\end{table*}

%% file: tables/ablation_table.tex
\begin{table}[ht]
\centering
\caption{Effectiveness of \our{} components with a large number of agents $m=64$. \textit{Base} is the model that only has improved communication.}
\label{table:ablation}
\resizebox{\columnwidth}{!}{%
\begin{tabular}{lccccc}
\toprule
 & \multicolumn{2}{c}{\texttt{den312d}} & \multicolumn{2}{c}{\texttt{warehouse}} \\
\cmidrule(lr){2-3} \cmidrule(lr){4-5}
Method & EL & SR & EL & SR \\
\midrule
Base & 140.70 & 97\% & 479.42 & 15\% \\
\quad + Prioritized Decisions & 129.84 & 100\% & 454.16 & 22\% \\
\quad\quad + Hybrid Guidance & 130.16 & 100\% & 255.31 & 88\% \\
\quad\quad\quad + Ensemble & 121.66 & 100\% & 189.58 & 100\% \\
\bottomrule
\end{tabular}
}
\end{table}

%% file: root.bbl
\begin{thebibliography}{10}
\providecommand{\url}[1]{#1}
\csname url@samestyle\endcsname
\providecommand{\newblock}{\relax}
\providecommand{\bibinfo}[2]{#2}
\providecommand{\BIBentrySTDinterwordspacing}{\spaceskip=0pt\relax}
\providecommand{\BIBentryALTinterwordstretchfactor}{4}
\providecommand{\BIBentryALTinterwordspacing}{\spaceskip=\fontdimen2\font plus
\BIBentryALTinterwordstretchfactor\fontdimen3\font minus \fontdimen4\font\relax}
\providecommand{\BIBforeignlanguage}[2]{{%
\expandafter\ifx\csname l@#1\endcsname\relax
\typeout{** WARNING: IEEEtran.bst: No hyphenation pattern has been}%
\typeout{** loaded for the language `#1'. Using the pattern for}%
\typeout{** the default language instead.}%
\else
\language=\csname l@#1\endcsname
\fi
#2}}
\providecommand{\BIBdecl}{\relax}
\BIBdecl

\bibitem{stern2019multi}
R.~Stern, N.~Sturtevant, A.~Felner, S.~Koenig, H.~Ma, T.~Walker, J.~Li, D.~Atzmon, L.~Cohen, T.~Kumar \emph{et~al.}, ``Multi-agent pathfinding: Definitions, variants, and benchmarks,'' in \emph{Proceedings of the International Symposium on Combinatorial Search}, vol.~10, no.~1, 2019, pp. 151--158.

\bibitem{wurman2008coordinating}
P.~R. Wurman, R.~D'Andrea, and M.~Mountz, ``Coordinating hundreds of cooperative, autonomous vehicles in warehouses,'' \emph{AI magazine}, vol.~29, no.~1, pp. 9--9, 2008.

\bibitem{morris2016planning}
R.~Morris, C.~S. Pasareanu, K.~S. Luckow, W.~Malik, H.~Ma, T.~S. Kumar, and S.~Koenig, ``Planning, scheduling and monitoring for airport surface operations.'' in \emph{AAAI Workshop: Planning for Hybrid Systems}, 2016, pp. 608--614.

\bibitem{ma2017feasibility}
H.~Ma, J.~Yang, L.~Cohen, T.~Kumar, and S.~Koenig, ``Feasibility study: Moving non-homogeneous teams in congested video game environments,'' in \emph{AAAI Conference on Artificial Intelligence and Interactive Digital Entertainment}, 2017.

\bibitem{yu2013structure}
J.~Yu and S.~LaValle, ``Structure and intractability of optimal multi-robot path planning on graphs,'' in \emph{AAAI Conference on Artificial Intelligence}, 2013.

\bibitem{banfi2017intractability}
J.~Banfi, N.~Basilico, and F.~Amigoni, ``Intractability of time-optimal multirobot path planning on 2d grid graphs with holes,'' \emph{IEEE Robotics and Automation Letters}, vol.~2, no.~4, pp. 1941--1947, 2017.

\bibitem{sharon2015conflict}
G.~Sharon, R.~Stern, A.~Felner, and N.~R. Sturtevant, ``Conflict-based search for optimal multi-agent pathfinding,'' \emph{Artificial Intelligence}, vol. 219, pp. 40--66, 2015.

\bibitem{barer2014suboptimal}
M.~Barer, G.~Sharon, R.~Stern, and A.~Felner, ``Suboptimal variants of the conflict-based search algorithm for the multi-agent pathfinding problem,'' in \emph{Proceedings of the International Symposium on Combinatorial Search}, vol.~5, no.~1, 2014, pp. 19--27.

\bibitem{li2021eecbs}
J.~Li, W.~Ruml, and S.~Koenig, ``Eecbs: A bounded-suboptimal search for multi-agent path finding,'' in \emph{Proceedings of the AAAI Conference on Artificial Intelligence}, vol.~35, no.~14, 2021, pp. 12\,353--12\,362.

\bibitem{li2021lifelong}
J.~Li, A.~Tinka, S.~Kiesel, J.~W. Durham, T.~S. Kumar, and S.~Koenig, ``Lifelong multi-agent path finding in large-scale warehouses,'' in \emph{Proceedings of the AAAI Conference on Artificial Intelligence}, vol.~35, no.~13, 2021, pp. 11\,272--11\,281.

\bibitem{sartoretti2019primal}
G.~Sartoretti, J.~Kerr, Y.~Shi, G.~Wagner, T.~S. Kumar, S.~Koenig, and H.~Choset, ``Primal: Pathfinding via reinforcement and imitation multi-agent learning,'' \emph{IEEE Robotics and Automation Letters}, 2019.

\bibitem{damani2021primal}
M.~Damani, Z.~Luo, E.~Wenzel, and G.~Sartoretti, ``Primal $ \_2 $: Pathfinding via reinforcement and imitation multi-agent learning-lifelong,'' \emph{IEEE Robotics and Automation Letters}, 2021.

\bibitem{ma2021learning}
Z.~Ma, Y.~Luo, and J.~Pan, ``Learning selective communication for multi-agent path finding,'' \emph{IEEE Robotics and Automation Letters}, vol.~7, no.~2, pp. 1455--1462, 2021.

\bibitem{ma2021distributed}
Z.~Ma, Y.~Luo, and H.~Ma, ``Distributed heuristic multi-agent path finding with communication,'' in \emph{2021 IEEE International Conference on Robotics and Automation (ICRA)}.\hskip 1em plus 0.5em minus 0.4em\relax IEEE, 2021, pp. 8699--8705.

\bibitem{mnih2013playing}
V.~Mnih, K.~Kavukcuoglu, D.~Silver, A.~Graves, I.~Antonoglou, D.~Wierstra, and M.~Riedmiller, ``Playing atari with deep reinforcement learning,'' \emph{arXiv preprint arXiv:1312.5602}, 2013.

\bibitem{wang2023scrimp}
Y.~Wang, B.~Xiang, S.~Huang, and G.~Sartoretti, ``{SCRIMP}: Scalable communication for reinforcement-and imitation-learning-based multi-agent pathfinding,'' \emph{arXiv preprint arXiv:2303.00605}, 2023.

\bibitem{lin2023sacha}
Q.~Lin and H.~Ma, ``{SACHA}: Soft actor-critic with heuristic-based attention for partially observable multi-agent path finding,'' \emph{IEEE Robotics and Automation Letters}, 2023.

\bibitem{gange2019lazy}
G.~Gange, D.~Harabor, and P.~J. Stuckey, ``Lazy cbs: implicit conflict-based search using lazy clause generation,'' in \emph{ICAPS}, 2019.

\bibitem{li2020new}
J.~Li, G.~Gange, D.~Harabor, P.~J. Stuckey, H.~Ma, and S.~Koenig, ``New techniques for pairwise symmetry breaking in multi-agent path finding,'' in \emph{Proceedings of the International Conference on Automated Planning and Scheduling}, vol.~30, 2020, pp. 193--201.

\bibitem{chung2023learning}
J.~Chung, J.~Fayyad, Y.~A. Younes, and H.~Najjaran, ``Learning to team-based navigation: A review of deep reinforcement learning techniques for multi-agent pathfinding,'' \emph{arXiv preprint arXiv:2308.05893}, 2023.

\bibitem{he2021asynchronous}
Z.~He, L.~Dong, C.~Sun, and J.~Wang, ``Asynchronous multithreading reinforcement-learning-based path planning and tracking for unmanned underwater vehicle,'' \emph{IEEE Transactions on Systems, Man, and Cybernetics: Systems}, vol.~52, no.~5, pp. 2757--2769, 2021.

\bibitem{wang2020mrcdrl}
D.~Wang, H.~Deng, and Z.~Pan, ``Mrcdrl: Multi-robot coordination with deep reinforcement learning,'' \emph{Neurocomputing}, 2020.

\bibitem{chen2022multiagent}
L.~Chen, Y.~Wang, Y.~Mo, Z.~Miao, H.~Wang, M.~Feng, and S.~Wang, ``Multiagent path finding using deep reinforcement learning coupled with hot supervision contrastive loss,'' \emph{IEEE Transactions on Industrial Electronics}, vol.~70, no.~7, pp. 7032--7040, 2022.

\bibitem{li2022multi}
W.~Li, H.~Chen, B.~Jin, W.~Tan, H.~Zha, and X.~Wang, ``Multi-agent path finding with prioritized communication learning,'' in \emph{2022 International Conference on Robotics and Automation (ICRA)}.\hskip 1em plus 0.5em minus 0.4em\relax IEEE, 2022, pp. 10\,695--10\,701.

\bibitem{chen2022multi}
L.~Chen, Y.~Wang, Z.~Miao, Y.~Mo, M.~Feng, and Z.~Zhou, ``Multi-agent path finding using imitation-reinforcement learning with transformer,'' in \emph{2022 IEEE International Conference on Robotics and Biomimetics (ROBIO)}.\hskip 1em plus 0.5em minus 0.4em\relax IEEE, 2022, pp. 445--450.

\bibitem{gao2023rde}
J.~Gao, Y.~Li, X.~Yang, and M.~Tan, ``Rde: A hybrid policy framework for multi-agent path finding problem,'' \emph{arXiv preprint arXiv:2311.01728}, 2023.

\bibitem{tang2024himap}
H.~Tang, F.~Berto, Z.~Ma, C.~Hua, K.~Ahn, and J.~Park, ``Hi{MAP}: Learning heuristics-informed policies for large-scale multi-agent pathfinding,'' in \emph{AAMAS}, 2024.

\bibitem{parisotto2020stabilizing}
E.~Parisotto, F.~Song, J.~Rae, R.~Pascanu, C.~Gulcehre, S.~Jayakumar, M.~Jaderberg, R.~L. Kaufman, A.~Clark, S.~Noury \emph{et~al.}, ``Stabilizing transformers for reinforcement learning,'' in \emph{International conference on machine learning}.\hskip 1em plus 0.5em minus 0.4em\relax PMLR, 2020, pp. 7487--7498.

\bibitem{vaswani2017attention}
A.~Vaswani, N.~Shazeer, N.~Parmar, J.~Uszkoreit, L.~Jones, A.~N. Gomez, {\L}.~Kaiser, and I.~Polosukhin, ``Attention is all you need,'' \emph{Advances in neural information processing systems}, vol.~30, 2017.

\bibitem{wang2016dueling}
Z.~Wang, T.~Schaul, M.~Hessel, H.~Hasselt, M.~Lanctot, and N.~Freitas, ``Dueling network architectures for deep reinforcement learning,'' in \emph{International conference on machine learning}.\hskip 1em plus 0.5em minus 0.4em\relax PMLR, 2016.

\bibitem{van2016deep}
H.~Van~Hasselt, A.~Guez, and D.~Silver, ``Deep reinforcement learning with double q-learning,'' in \emph{AAAI conference on artificial intelligence}, 2016.

\bibitem{hasselt2010double}
H.~Hasselt, ``Double q-learning,'' \emph{Advances in neural information processing systems}, vol.~23, 2010.

\bibitem{latombe2012robot}
J.-C. Latombe, \emph{Robot motion planning}.\hskip 1em plus 0.5em minus 0.4em\relax Springer Science \& Business Media, 2012, vol. 124.

\bibitem{silver2005cooperative}
D.~Silver, ``Cooperative pathfinding,'' in \emph{AAAI conference on artificial intelligence and interactive digital entertainment}, 2005.

\bibitem{okumura2022priority}
K.~Okumura, M.~Machida, X.~D{\'e}fago, and Y.~Tamura, ``Priority inheritance with backtracking for iterative multi-agent path finding,'' \emph{Artificial Intelligence}, vol. 310, p. 103752, 2022.

\bibitem{ma2019searching}
H.~Ma, D.~Harabor, P.~J. Stuckey, J.~Li, and S.~Koenig, ``Searching with consistent prioritization for multi-agent path finding,'' in \emph{AAAI conference on artificial intelligence}, 2019.

\bibitem{schaul2015prioritized}
T.~Schaul, J.~Quan, I.~Antonoglou, and D.~Silver, ``Prioritized experience replay,'' \emph{arXiv preprint arXiv:1511.05952}, 2015.

\bibitem{bengio2009curriculum}
Y.~Bengio, J.~Louradour, R.~Collobert, and J.~Weston, ``Curriculum learning,'' in \emph{Proceedings of the 26th annual international conference on machine learning}, 2009, pp. 41--48.

\bibitem{schulman2017proximal}
J.~Schulman, F.~Wolski, P.~Dhariwal, A.~Radford, and O.~Klimov, ``Proximal policy optimization algorithms,'' \emph{arXiv preprint arXiv:1707.06347}, 2017.

\bibitem{ferner2013odrm}
C.~Ferner, G.~Wagner, and H.~Choset, ``Odrm* optimal multirobot path planning in low dimensional search spaces,'' in \emph{ICRA}.\hskip 1em plus 0.5em minus 0.4em\relax IEEE, 2013.

\bibitem{williams1992simple}
R.~J. Williams, ``Simple statistical gradient-following algorithms for connectionist reinforcement learning,'' \emph{Machine learning}, 1992.

\bibitem{berto2024rl4co}
F.~Berto, C.~Hua, J.~Park, L.~Luttmann, Y.~Ma, F.~Bu, J.~Wang, H.~Ye, M.~Kim, S.~Choi, N.~G. Zepeda, A.~Hottung, J.~Zhou, J.~Bi, Y.~Hu, F.~Liu, H.~Kim, J.~Son, H.~Kim, D.~Angioni, W.~Kool, Z.~Cao, J.~Zhang, K.~Shin, C.~Wu, S.~Ahn, G.~Song, C.~Kwon, L.~Xie, and J.~Park, ``{RL4CO: an Extensive Reinforcement Learning for Combinatorial Optimization Benchmark},'' \emph{arXiv preprint arXiv:2306.17100}, 2024, \url{https://github.com/ai4co/rl4co}.

\bibitem{hottung2019neural}
A.~Hottung and K.~Tierney, ``Neural large neighborhood search for the capacitated vehicle routing problem,'' \emph{arXiv preprint arXiv:1911.09539}, 2019.

\bibitem{kool2022euro}
W.~Kool, L.~Bliek, D.~Numeroso, Y.~Zhang, T.~Catshoek, K.~Tierney, T.~Vidal, and J.~Gromicho, ``The euro meets neurips 2022 vehicle routing competition,'' in \emph{NeurIPS 2022 Competition Track}.\hskip 1em plus 0.5em minus 0.4em\relax PMLR, 2022, pp. 35--49.

\bibitem{ye2023glop}
H.~Ye, J.~Wang, H.~Liang, Z.~Cao, Y.~Li, and F.~Li, ``Glop: Learning global partition and local construction for solving large-scale routing problems in real-time,'' \emph{AAAI 2024}, 2024.

\bibitem{ye2024deepaco}
H.~Ye, J.~Wang, Z.~Cao, H.~Liang, and Y.~Li, ``Deepaco: Neural-enhanced ant systems for combinatorial optimization,'' \emph{Advances in Neural Information Processing Systems}, vol.~36, 2024.

\bibitem{liu2024example}
F.~Liu, X.~Tong, M.~Yuan, X.~Lin, F.~Luo, Z.~Wang, Z.~Lu, and Q.~Zhang, ``An example of evolutionary computation+ large language model beating human: Design of efficient guided local search,'' \emph{arXiv preprint arXiv:2401.02051}, 2024.

\bibitem{ye2024reevo}
H.~Ye, J.~Wang, Z.~Cao, F.~Berto, C.~Hua, H.~Kim, J.~Park, and G.~Song, ``Large language models as hyper-heuristics for combinatorial optimization,'' \emph{arXiv preprint arXiv:2402.01145}, 2024, \url{https://github.com/ai4co/LLM-as-HH}.

\end{thebibliography}
